\documentstyle[12pt,dina4p,epsfig]{article}

\def\Journal#1#2#3#4{{#1} {\bf #2}, #3 (#4)}

\def\NPB{{\em Nucl. Phys.} B}

\def\PRL{\em Phys. Rev. Lett.}
\def\PRD{{\em Phys. Rev.} D}
\def\ZPC{{\em Z. Phys.} C}

\def\etjet{E_T^{jet}}
\def\etajet{\eta^{jet}}

\def\etaphi{\eta-\varphi}
 
\def\rs{R_{SEP}}
 
\def\etar{-1<\etajet<2}

\def\q2{Q^2}
\def\pb1{pb$^{-1}$}

\def\g2{GeV$^2$}
 
\def\rr1{R=1}
\def\r7{R=0.7}
\def\R71{R=0.7\ {\rm and}\ 1}

\def\mj{M^{JJ}}
\def\smj{d\sigma/d\mj}
\def\cost{\vert\cos\theta^*\vert}
\def\scost{d\sigma/d\cost}

\begin{document}
 
\title{Jet Production at HERA$^\dag$
\footnotetext{$^\dag$To appear in the Proccedings of the ``International
Europhysics Conference on High Energy Physics'', Jerusalem, Israel, August
1997.}}
 
\author{C. Glasman\\
(representing the H1 and ZEUS Collaborations)\\
DESY, Germany}
\date{ }

\maketitle

\begin{abstract}
Studies on the structure of the photon are presented by means of the extraction
of a leading order effective parton distribution in the photon and measurements
of inclusive jet differential cross sections in photoproduction. Measurements
of the internal structure of jets have been performed and are also presented
as a function of the transverse energy and pseudorapidity of the jets. Dijet
cross sections have been measured as a function of the dijet mass and
centre-of-mass scattering angle.
\end{abstract}

The main source of jets at HERA\footnote{HERA provides collisions between
positrons of energy $E_e=27.5$ GeV and protons of energy $E_p=820$ GeV.} comes
from collisions between protons and quasi-real photons ($\q2\approx 0$, where
$\q2$ is the virtuality of the photon) emitted by the positron beam
(photoproduction). At lowest order QCD, two hard scattering processes
contribute to jet production: the resolved process, in which a parton from the
photon interacts with a parton from the proton, producing two jets in the
final state; and the direct process, in which the photon interacts pointlike
with a parton from the proton, also producing two jets in the final state.

The study of high-$p_T$ jet photoproduction provides tests of QCD and allows
to probe the structure of the photon. In perturbative QCD the cross section
for jet production is given by

\begin{equation}
{d^4\sigma\over dy dx_{\gamma} dx_p d\cos\theta^*}\propto
{f_{\gamma /e}(y)\over y}
\sum_{ij} {f_{i/\gamma}(x_{\gamma},\hat\q2)\over 
x_{\gamma}}{f_{j/p}(x_p,\hat\q2)\over x_p}
\vert M_{ij}(\cos\theta^*)\vert^2
\label{eq:one}
\end{equation}
where $f_{\gamma /e}(y)$ is the flux of photons from the positron
approximated by the Weizs\"{a}cker-Williams formula; $y=E_{\gamma}/E_e$ is
the inelasticity parameter; $f_{i/\gamma}(x_{\gamma},\hat\q2)\ 
(f_{j/p}(x_p,\hat\q2))$ are the parton
densities in the photon\footnote{The resolved and direct processes are
included in $f_{i/\gamma}$.} (proton) at a scale $\hat Q$;
$x_{\gamma}=E_i/E_{\gamma}\ (x_p=E_j/E_p)$ is the fractional momentum of
the incoming parton from the photon (proton); and
$\vert M_{ij}(\cos\theta^*)\vert^2$ are the QCD matrix elements for the
parton-parton scattering.

Two approaches to the study of the structure of the photon are presented here.
One approach is to extract directly from the data an effective parton
distribution in the photon. The second approach is to measure jet cross
sections that can be calculated theoretically which then provide a testing
ground for parametrisations of the photon structure function.

The method \cite{p:comb} to extract an effective parton distribution in the
photon is based on the use of the leading order (LO) matrix elements for the
subprocesses with gluon exchange, which give the dominant contribution to the
jet cross section in resolved processes for the kinematic regime studied.
Then, the quark and gluon densities are combined into an effective parton
distribution:

\begin{equation}
\sum [ f_{q/{\gamma}}(x_{\gamma},p_T^2)+
f_{\bar q/{\gamma}}(x_{\gamma},p_T^2) ] +
{9\over 4}f_{g/{\gamma}}(x_{\gamma},p_T^2)
\label{eq:two}
\end{equation}

\begin{figure}
\centerline{\mbox{
\epsfig{figure=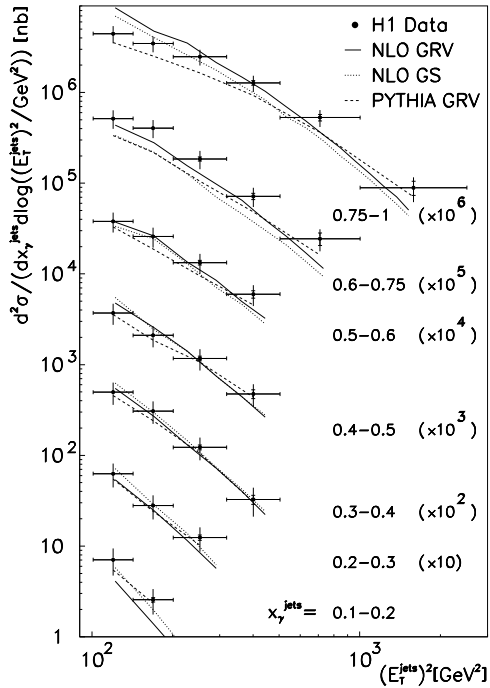,width=7cm}}}
\caption{Double differential dijet cross sections.
\label{fig:cero}}
\end{figure}

\begin{figure}
\centerline{\mbox{
\epsfig{figure=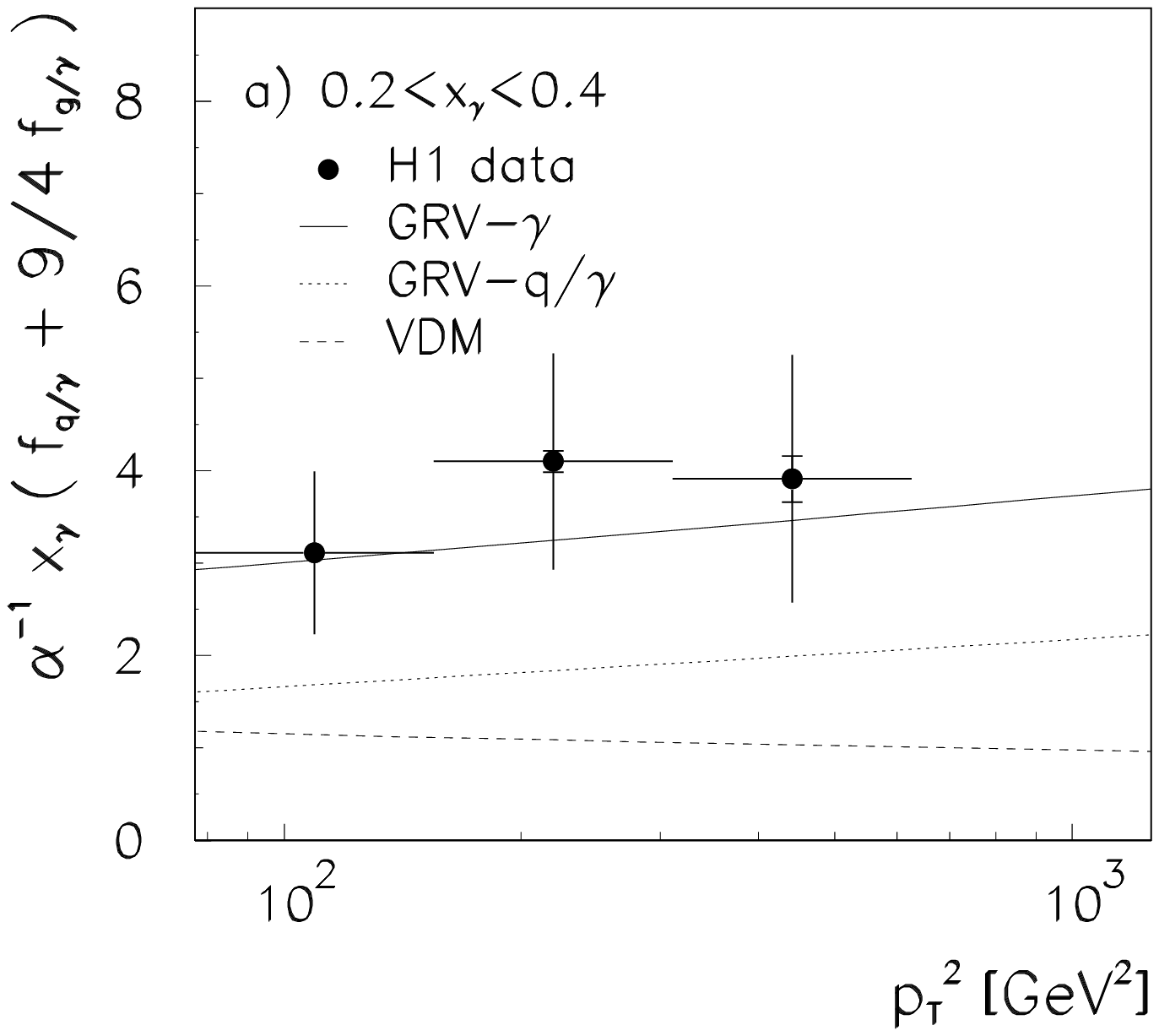,width=7cm}
\epsfig{figure=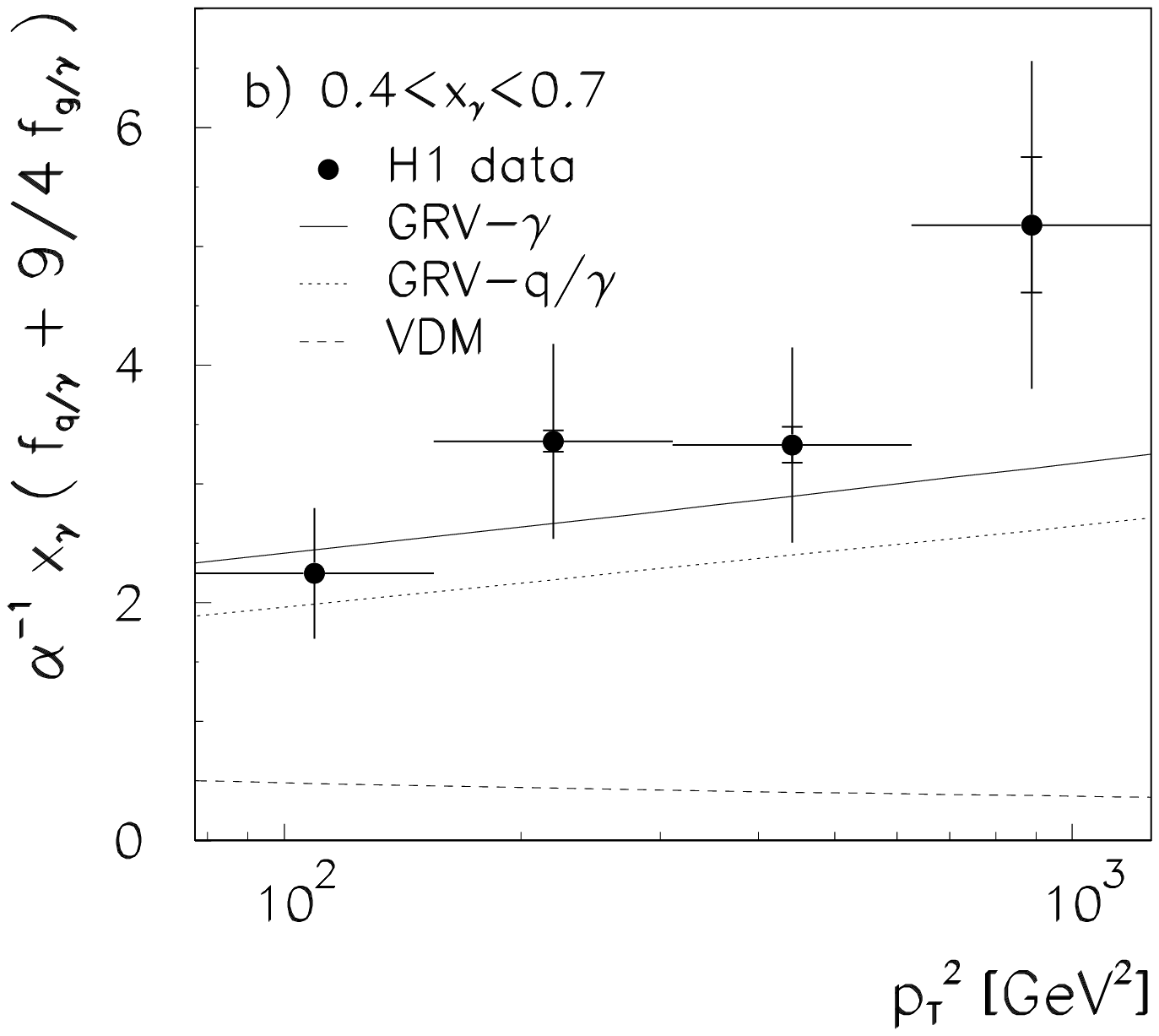,width=7cm}}}
\caption{Leading order effective parton distribution in the photon.
\label{fig:one}}
\end{figure}

A LO effective parton distribution in the photon was extracted by unfolding
the double differential dijet cross section as a function of the average
transverse energy of the two jets with highest transverse energy and of the
fraction of the photon's energy ($x_{\gamma}$) participating in the production
of the two highest-transverse-energy jets (see figure~\ref{fig:cero}). The
measurement was performed using a fixed cone algorithm \cite{p:cone} with
radius $\r7$ in the pseudorapidity\footnote{The pseudorapidity is defined as
$\eta=-\ln(\tan\frac{\theta}{2})$, where the polar angle $\theta$ is taken
with respect to the proton beam direction.} ($\eta$) $-$ azimuth ($\varphi$)
plane and in the kinematic range $0.2<y<0.83$ and $\q2<4$ \g2.
Figure~\ref{fig:one} shows the extracted effective parton distribution as a
function of the scale $p_T$ (the transverse momentum of the parton) for two
ranges of $x_{\gamma}$. The data exhibit an increase with the scale $p_T$
which is compatible with the logarithmic increase predicted by QCD (i.e., the
anomalous component of the photon structure function). The predictions using
the GRV-LO \cite{p:grvlo} parametrisations of the parton densities in the
photon and the pion are also shown in figure~\ref{fig:one}. The data
disfavour a purely hadronic behaviour and are compatible with the prediction
which includes the anomalous component.

The second approach to the study of the photon structure function is to provide
measurements of differential jet cross sections at high jet transverse
energies, where the proton parton densities are constrained by other
measurements, and, therefore, are sensitive to the photon structure function.

\begin{figure}
\centerline{\mbox{
\epsfig{figure=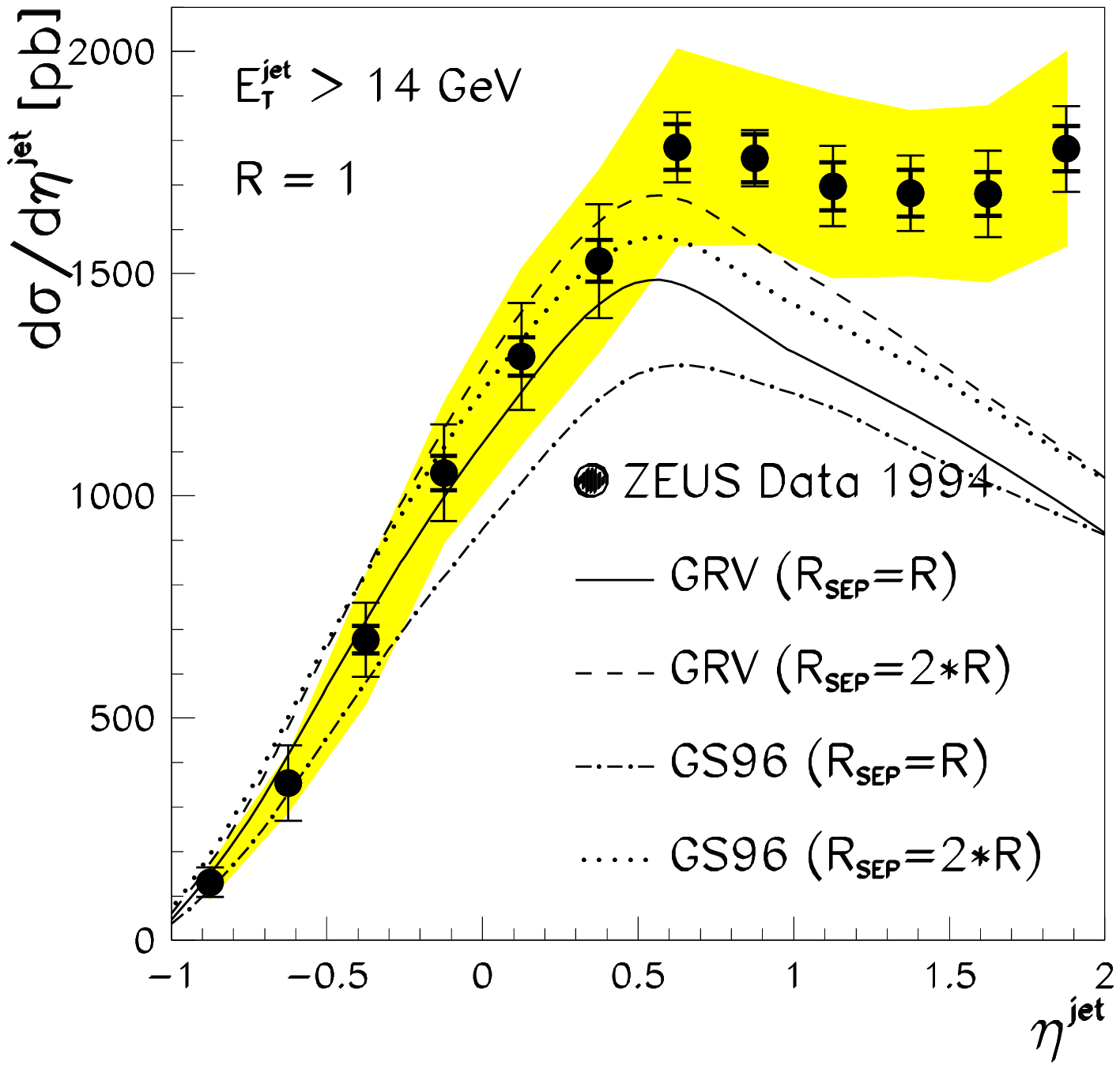,width=8cm}
\hspace{-1cm}
\epsfig{figure=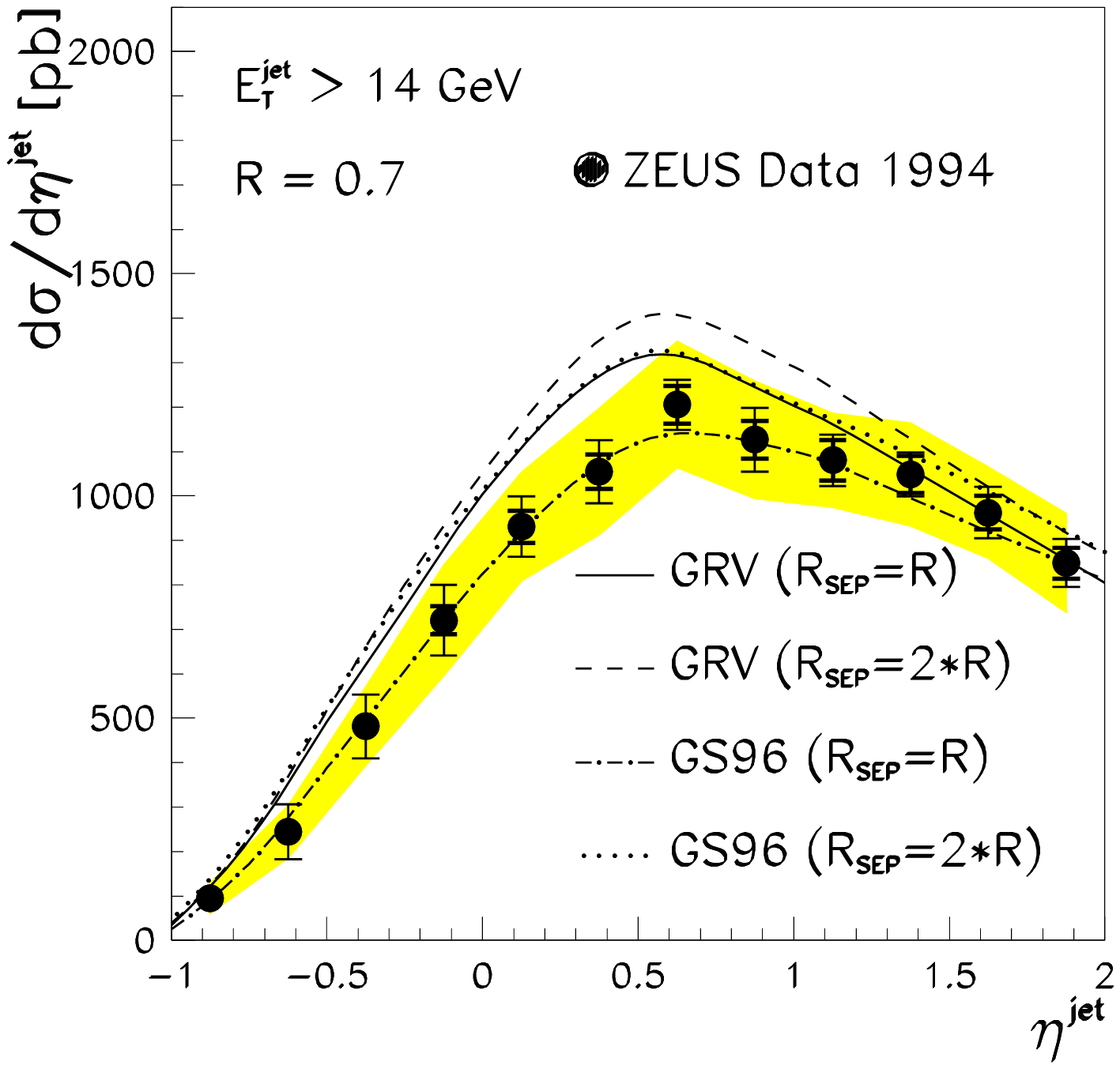,width=8cm}}}
\vspace{-1cm}
\caption{Inclusive jet differential cross sections.
\label{fig:two}}
\end{figure}

Figure~\ref{fig:two} shows the measurements of the inclusive jet differential
cross sections for jets searched with an iterative cone algorithm
\cite{p:cone} as a function of the jet pseudorapidity ($\etajet$) for jets
with transverse energy satisfying $\etjet>14$ GeV. The measurements were
performed in the kinematic region given by $0.2<y<0.85$ and $\q2\leq 4$ \g2,
and for two cone radii: $\R71$. The behaviour of the cross section as a
function of $\etajet$ in the region $\etajet>1$ is very different for the two
cone radii: it is flat for $\rr1$ whereas it decreases as $\etajet$ increases
for $\r7$.

Next-to-leading order (NLO) QCD calculations \cite{p:kramer3} are compared to
the measurements in figure~\ref{fig:two} using two different parametrisations
of the photon structure function: GRV-HO \cite{p:grvho} and GS96
\cite{p:gs}, and two values of\footnote{The parameter $\rs$ is introduced
into the NLO calculations in order to simulate the experimental jet algorithm
by adjusting the minimum distance in $\etaphi$ at which two partons are no
longer merged into a single jet \cite{p:sdellis}.} $\rs$. The CTEQ4M
\cite{p:cteq4} proton parton densities have been used in all cases. For
forward jets with $\rr1$ an excess of the measurements with respect to the
calculations is observed. This discrepancy is attributed to a possible
contribution from non-perturbative effects (e.g., the so-called ``underlying
event''), which are not included in the theoretical calculations. This
contribution is supposed to be reduced by decreasing the size of the cone
since the transverse energy density inside the cone of the jet due to the
underlying event is expected to be roughly proportional to the area covered by
the cone. Good agreement between data and NLO calculations is observed for
measurements performed using a cone radius of $\r7$ for the entire $\etajet$
range measured. The predictions using GRV-HO and GS96 show differences which
are of the order of the largest systematic uncertainty of the measurements.
Thus, these measurements exhibit a sensitivity to the parton densities in the
photon and can be used in quantitative studies.

To study the internal structure of the jets, the jet shape $\psi(r)$ has been
used. $\psi(r)$ is defined as the average fraction of the jet's transverse
energy that lies inside an inner cone of radius $r$, concentric with the jet
defining cone~\cite{p:sdellis}:

\begin{equation}
\psi(r) = \frac{1}{N_{jets}} \sum_{jets} \frac{E_T(r)}{E_T(r=R)}
\label{eq:three}
\end{equation}
where $E_T(r)$ is the transverse energy within the inner cone and $N_{jets}$
is the total number of jets in the sample. By definition, $\psi(r=R)=1$. The
jet shape is determined by fragmentation and gluon radiation. However, at
sufficiently high $\etjet$ the most important contribution is predicted to
come from gluon emission off the primary parton.

\begin{figure}
\centerline{\mbox{
\epsfig{figure=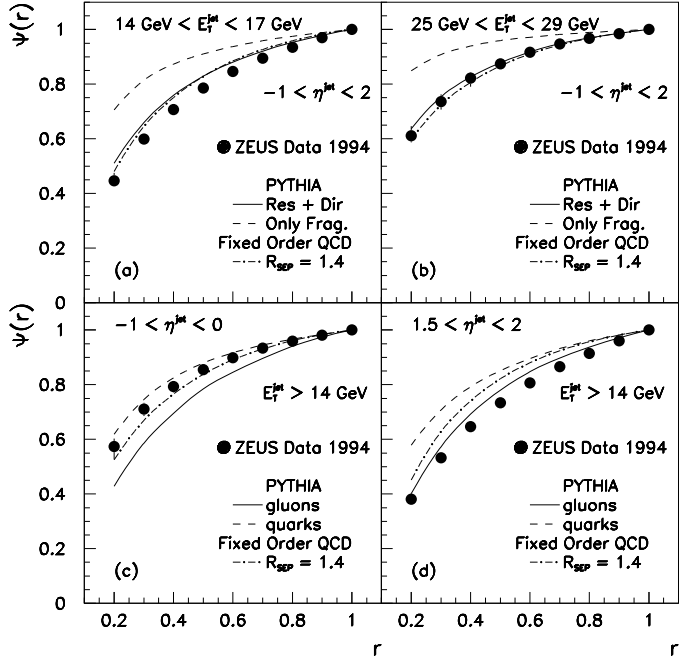,width=10cm}}}
\caption{Jet shapes: $\etjet$ and $\etajet$ dependence.
\label{fig:three}}
\end{figure}

Figures~\ref{fig:three}a and \ref{fig:three}b show the measured jet shapes
$\psi(r)$ as a function of the inner cone radius $r$ using a cone algorithm
with radius $\rr1$ for jets with $\etar$ and in two regions of $\etjet$. As a
jet becomes narrower, the value of $\psi(r)$ increases for a fixed value of
$r$. It is observed that the jets become narrower as $\etjet$ increases. For
comparison, the predictions from leading-logarithm parton-shower Monte Carlo
calculations as implemented in the PYTHIA generator for resolved plus direct
processes are shown. The predictions reproduce reasonably well the data except
in the lowest-$\etjet$ region where small differences between data and the
predictions are observed. PYTHIA including resolved plus direct processes but
without initial and final state parton radiation predicts jet shapes which are
too narrow in each region of $\etjet$. These comparisons show that parton
radiation is the dominant mechanism responsible for the jet shape in the range
of $\etjet$ studied.

The $\etajet$ dependence of the jet shape is presented in
figures~\ref{fig:three}c and \ref{fig:three}d. It is observed that the jets
become broader as $\etajet$ increases. Perturbative QCD predicts that gluon
jets are broader than quark jets as a consequence of the fact that the
gluon-gluon is larger than the quark-gluon coupling strength. The predictions
of PYTHIA for quark and gluon jets are also shown. The data go from being
dominated by quark jets in the final state ($\etajet<0$) to being dominated by
gluon jets ($\etajet>1.5$). Therefore, the broadening of the measured jet
shapes as $\etajet$ increases is consistent with an increase of the fraction
of gluon jets.

Lowest non-trivial-order QCD calculations \cite{p:kramer} of the jet shapes
are compared to the measurements in figure~\ref{fig:three}. The fixed-order
QCD calculations with a common value of $\rs=1.4$ reproduce reasonably well
the measured jet shapes in the region $\etjet>17$ GeV and in the region
$-1<\etajet<1$.

\begin{figure}
\centerline{\mbox{
\epsfig{figure=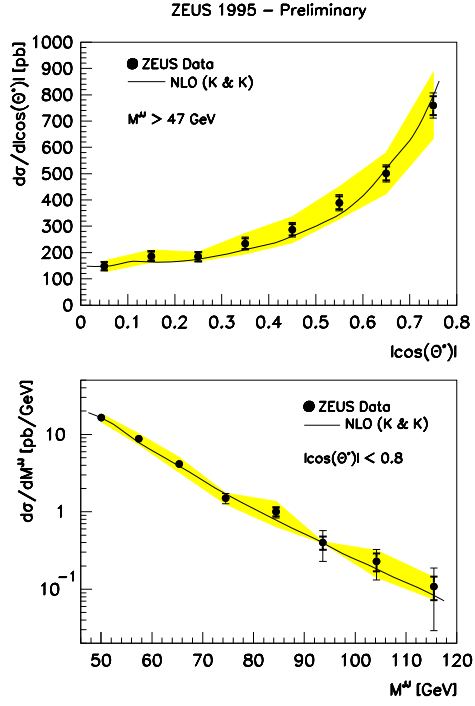,width=10cm}}}
\caption{Dijet differential cross sections.
\label{fig:five}}
\end{figure}

The dijet mass ($\mj$) distribution provides a test of QCD and is sensitive
to the presence of new particles or resonances that decay into two jets.
The dijet cross section as a function of the scattering angle in the dijet
centre-of-mass system ($\cos\theta^*$) reflects the underlying parton dynamics.
New particles or resonances decaying into two jets may also be identified by
deviations in the $\cost$ distribution with respect to the predictions.
$\smj$ has been measured for $\cost <0.8$, and $\scost$ has been measured for
$\mj >47$~GeV. The results are presented in figure~\ref{fig:five}. The
measured $\scost$ increases as $\cost$ increases. The measured $\smj$
exhibits a steep fall-off of 2 orders of magnitude in the $\mj$ range
considered. NLO QCD calculations \cite{p:kramer3} are compared to the
measurements in figure~\ref{fig:five}. The CTEQ4M (GS96) parametrisations of
the proton (photon) parton densities have been used. The prediction for
$\scost$, which is normalised to the lowest-$\cost$ data point, agrees in
shape reasonably well with the measured distribution. The prediction for
$\smj$ describes the shape and magnitude of the measured distribution up to
the highest $\mj$ studied ($\sim 120$ GeV).

The results on photoproduction of jets presented here constitute a
step forward towards testing QCD and the extraction of the photon parton
densities.

\end{document}